\documentclass[sigconf,nonacm]{acmart}

\usepackage{graphicx}  
\usepackage{booktabs}  
\usepackage{amsmath}   
\usepackage{hyperref}  
\usepackage{multirow}  
\usepackage{balance}   
\usepackage[english]{babel}
\usepackage{microtype}  
\usepackage{tabularx}
\usepackage[justification=centering]{caption}
\usepackage{subcaption}  
\usepackage{enumitem}

\title{Detecting Fake News Belief via Skin and Blood Flow Signals
}

\author{Gennie Nguyen}
\affiliation{%
  \institution{The Australian National University}
  \city{Canberra}
  \state{ACT}
  \country{Australia}
}
\email{Gennie.Nguyen@anu.edu.au}

\author{Lei Wang}
\affiliation{%
  \institution{Griffith University}
  \city{Brisbane}
  \state{Queensland}
  \country{Australia}
}
\email{l.wang4@griffith.edu.au}

\author{Yangxueqing Jiang}
\affiliation{%
  \institution{School of Medicine and Psychology, The Australian National University}
  \city{Canberra}
  \state{ACT}
  \country{Australia}
}
\email{Yangxueqing.jiang@anu.edu.au}

\author{Tom Gedeon}
\affiliation{%
  \institution{Optus Centre for AI, Curtin University}
  \city{Perth}
  \state{Western Australia}
  \country{Australia}
}
\email{tom.gedeon@curtin.edu.au}

\date{February 2025}

\makeatletter
\DeclareRobustCommand\onedot{\futurelet\@let@token\bmv@onedotaux}
\def\bmv@onedotaux{\ifx\@let@token.\else.\null\fi\xspace}
\def\eg{\emph{e.g}.}

 \def\vs{\emph{vs}.}

\makeatother

\begin{document}
\settopmatter{printfolios=true}

\begin{abstract}

Misinformation poses significant risks to public opinion, health, and security. While most fake news detection methods rely on text analysis, little is known about how people physically respond to false information or repeated exposure to the same statements. This study investigates whether wearable sensors can detect belief in a statement or prior exposure to it.
We conducted a controlled experiment where participants evaluated statements while wearing an EmotiBit sensor that measured their skin conductance (electrodermal activity, EDA) and peripheral blood flow (photoplethysmography, PPG). From 28 participants, we collected a dataset of 672 trials, each labeled with whether the participant believed the statement and whether they had seen it before. This dataset introduces a new resource for studying physiological responses to misinformation.
Using machine learning models, including KNN, CNN, and LightGBM, we analyzed these physiological patterns. The best-performing model achieved 67.83\% accuracy, with skin conductance outperforming PPG. These findings demonstrate the potential of wearable sensors as a minimally intrusive tool for detecting belief and prior exposure, offering new directions for real-time misinformation detection and adaptive, user-aware systems.



\end{abstract}

\maketitle

\section{Introduction}

The rapid advancement of digital technologies and the widespread adoption of social media have dramatically transformed how we consume and share information \cite{wang2025time}. While these platforms offer unprecedented convenience and access to knowledge, they also enable the swift dissemination of misinformation --- posing significant risks to public opinion, health, and national security. From misleading headlines to fabricated statistics, fake news can influence beliefs, decisions, and behavior on a large scale.

Most existing fake news detection research has focused on analyzing the content of news articles using natural language processing (NLP) techniques, yielding promising results in text-based classification~\cite{saleh2024comprehensive, de2021identifying}. However, relatively little is known about how individuals physiologically respond to false information or repeated exposure to the same statements. Human reactions to misinformation are not purely cognitive --- they also involve conscious and involuntary emotional and physiological processes. Understanding these responses can provide valuable insights into how people internalize, reject, or react to repeated exposure to misleading content, even when strong familiarity has not yet formed.


\begin{figure}[tbp]
\centering
\begin{subfigure}[b]{0.48\linewidth}
    \centering
    \includegraphics[width=\linewidth]{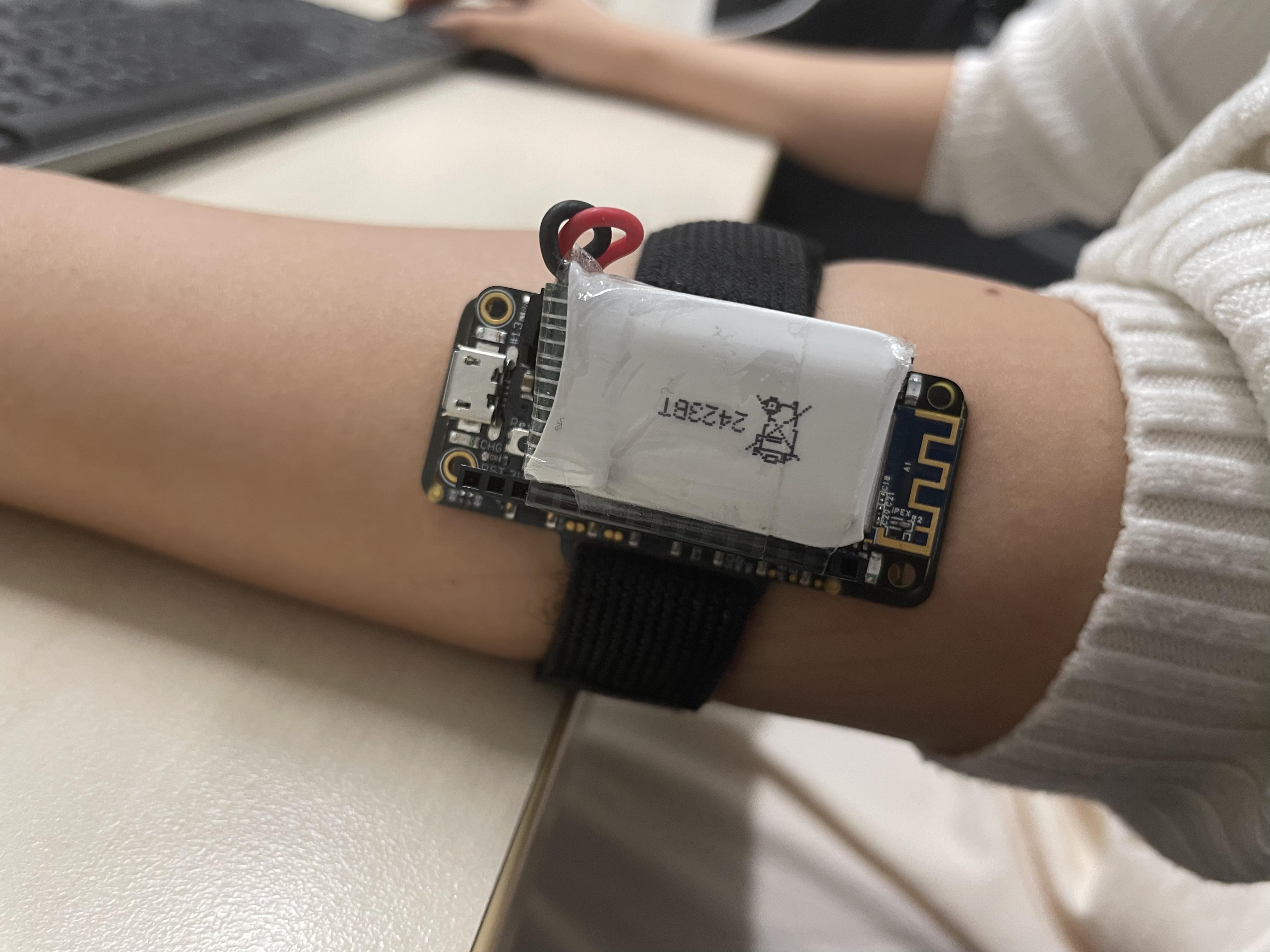}
    \caption{}
    \label{fig:a}
\end{subfigure}
\hfill
\begin{subfigure}[b]{0.48\linewidth}
    \centering
    \includegraphics[width=\linewidth]{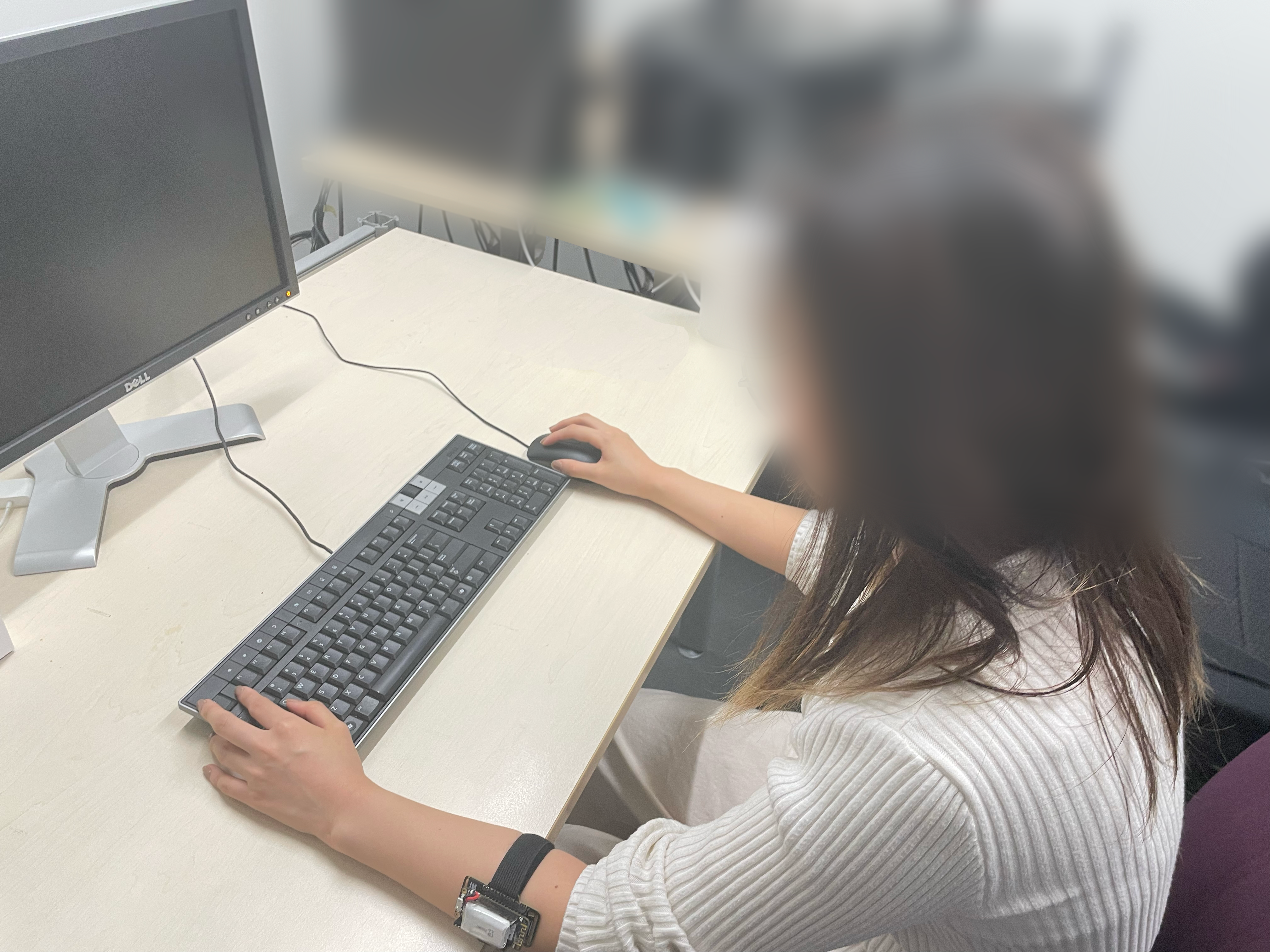}
    \caption{}
    \label{fig:b}
\end{subfigure}
\vspace{1em}
\begin{subfigure}[b]{\columnwidth}
    \centering
    \includegraphics[trim=0.4cm 0.5cm 0.4cm 0cm, clip=true, width=\columnwidth]{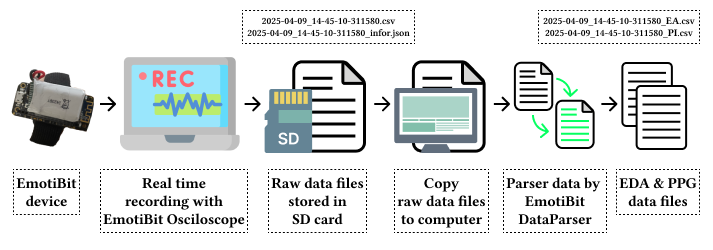}
    \caption{}
    \label{fig:c}
\end{subfigure}
\caption{Experimental setup for physiological signal collection using the EmotiBit sensor.
(a) Close-up of the EmotiBit wearable sensor attached to the participant's forearm (b) A participant (c) Data acquisition workflow.
}
\label{experiment_setup}
\end{figure}

In cognitive and affective science, physiological signals such as electrodermal activity (EDA) and photoplethysmography (PPG) have proven effective in capturing emotional arousal and cognitive load. EDA measures skin conductance, reflecting sympathetic nervous system activity and emotional reactivity~\cite{caruelle2019use}, while PPG captures heart rate variability (HRV), often linked to mental effort and stress~\cite{allen2021photoplethysmography}. Unlike brain-signal measurements like EEG, which are expensive, obtrusive, and difficult to scale, EDA and PPG can be captured conveniently using affordable, wearable sensors. Despite their potential, these signals remain underexplored in the context of belief in misinformation and responses to repeated claims.


Our study addresses this gap by introducing a novel dataset and empirical framework that connect physiological signals with belief and familiarity judgments. Specifically, we conducted a controlled experiment in which participants evaluated informational claims while wearing an EmotiBit sensor (see Fig.~\ref{experiment_setup}) that recorded EDA and PPG signals. Each trial was annotated based on whether the participant believed the statement and whether they had seen it before. This dataset provides a new resource for studying the physiological correlates of misinformation processing.

In this work, we pose two core research questions: (i) \textbf{Do physiological signals such as EDA and PPG reflect belief and familiarity with misinformation?} That is, can these signals serve as markers of the emotional and cognitive processes underlying judgments about false or repeated information? (ii) \textbf{Can machine learning models accurately classify belief and repetition status from physiological data?} If so, which signal type and model architecture offer the best predictive performance?

To answer these questions, we trained and evaluated several machine learning models, including K-Nearest Neighbors (KNN), Convolutional Neural Networks (CNN), and Light Gradient Boosting Machines (LightGBM), to classify participants’ belief and repetition responses based on physiological features. Our comparative analysis shows that skin conductance (EDA) outperforms blood flow (PPG) in predictive accuracy, with the best-performing model achieving 67.83\% accuracy. In summary, this paper contributes:
\renewcommand{\labelenumi}{\roman{enumi}.}
\begin{enumerate}[leftmargin=0.6cm]
\item A \textbf{new dataset} linking wearable-recorded physiological signals (EDA and PPG) with human judgments about belief and repetition in misinformation, addressing a critical gap in the literature;
\item An \textbf{empirical benchmark} evaluating multiple machine learning models and signal types for classifying responses to fake news;
\item \textbf{Evidence for the potential} of wearable biosignals as a minimally invasive approach to belief modeling, opening new directions for adaptive, user-aware misinformation detection systems.
\end{enumerate}

\begin{figure}[tbp]
    \centering
    \includegraphics[width=\columnwidth]{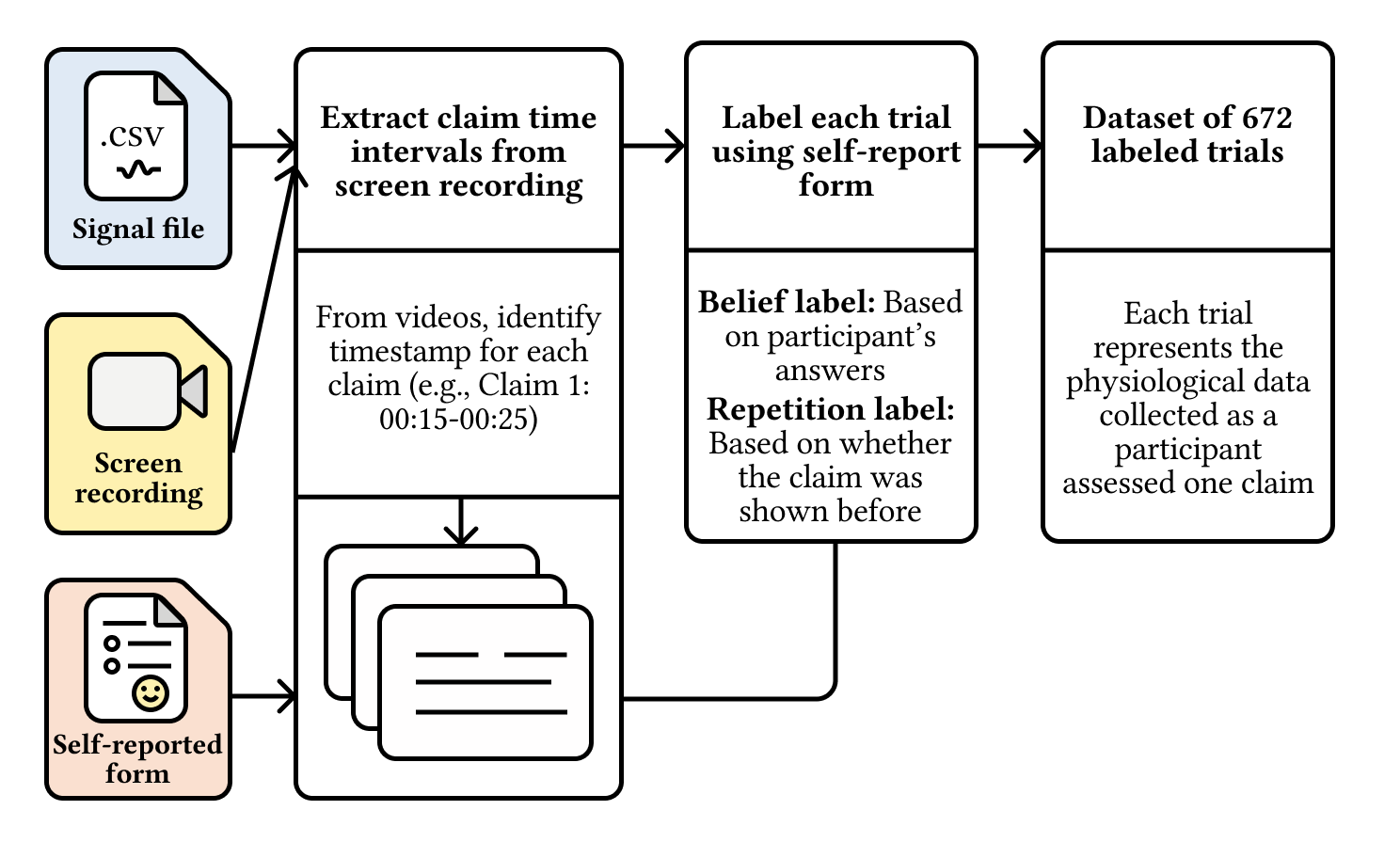}
    \caption{Data annotation process.
    Raw EDA/PPG signals, screen recordings, and participant self-reports were synchronized based on timestamps to align with each claim. These synchronized intervals were then used to segment the EDA data and assign belief and repetition labels to each segment, creating a structured dataset at the trial level.
    }
    \label{fig:data_align}
\end{figure}

\section{Related Work}

\begin{figure}[tbp]
\centering
\begin{subfigure}[b]{0.495\linewidth}
\centering\includegraphics[trim=0 0 0 0, clip=true,width=\linewidth]{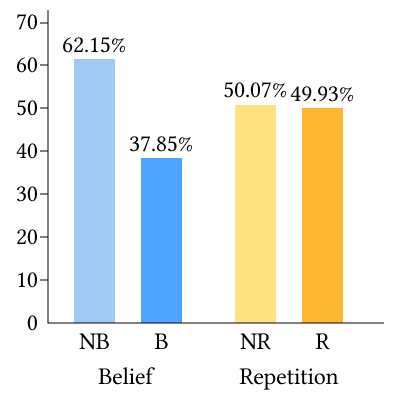}
\caption{\label{fig:belief_repetion}}
\end{subfigure}
\begin{subfigure}[b]{0.495\linewidth}
\centering\includegraphics[trim=0 0 0 0, clip=true,width=\linewidth]{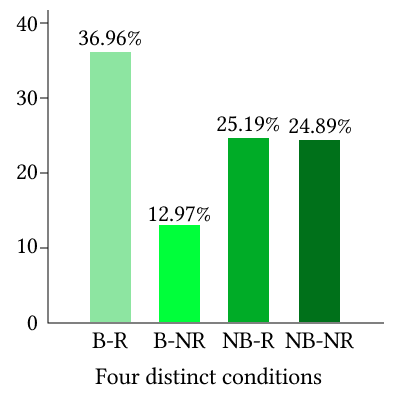}
\caption{\label{fig:four_distinct}}
\end{subfigure}
\caption{Label \% distribution across the dataset.
(a) Distribution of the Belief and Repetition labels. The dataset has a slight skew to the Not Believed class, Repetition is nearly balanced.
(b) Distribution across four combined conditions: Believed and Repeated (B-R), Believed and Not Repeated (B-NR), Not Believed and Repeated (NB-R), and Not Believed and Not Repeated (NB-NR).
}
\label{fig:label_dist}
\end{figure}

\begin{figure*}[tbp]
    \centering
    \begin{subfigure}[t]{\textwidth}
        \centering
        \begin{subfigure}[t]{0.32\textwidth}
            \includegraphics[width=\textwidth]{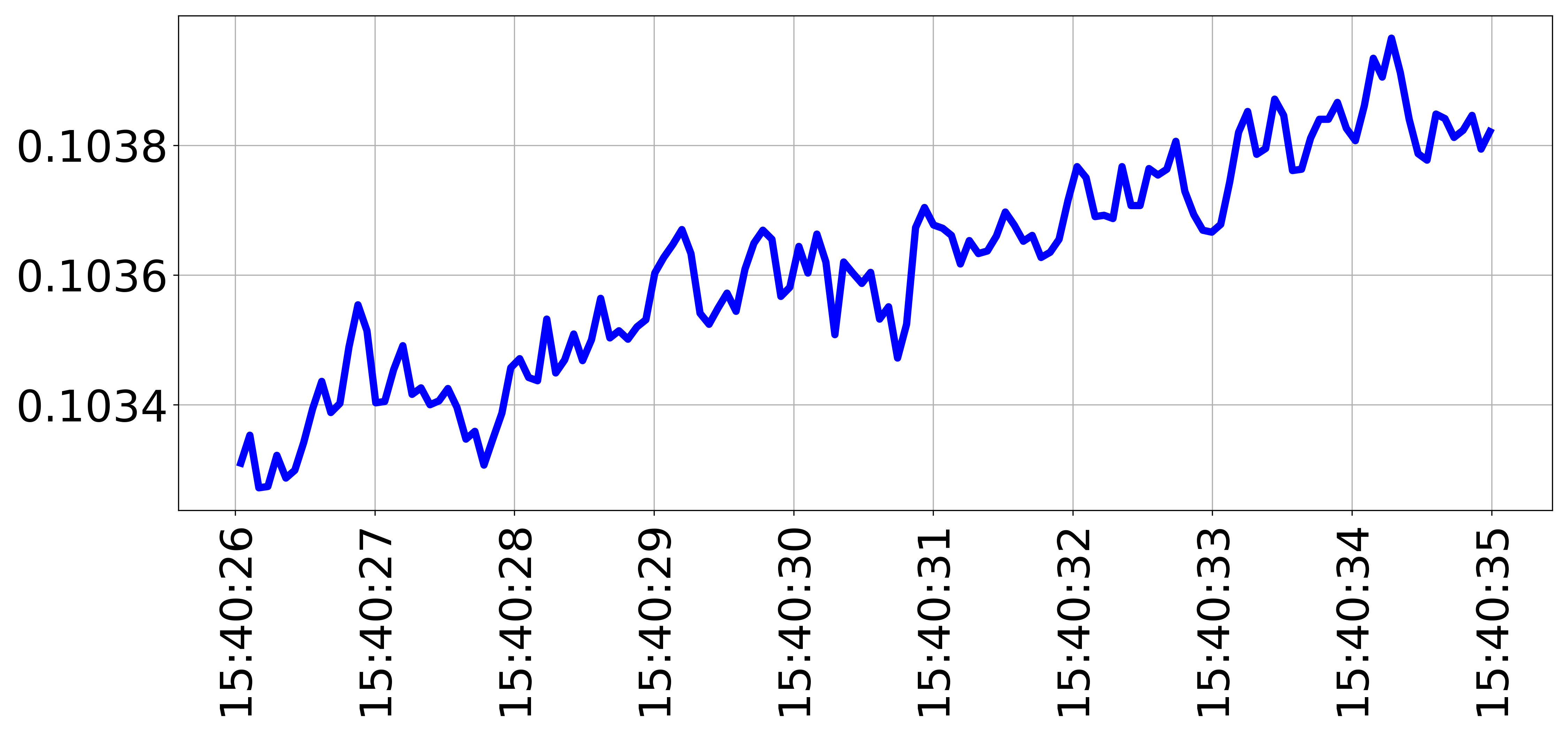}
        \end{subfigure}
        \hfill
        \begin{subfigure}[t]{0.32\textwidth}
            \includegraphics[width=\textwidth]{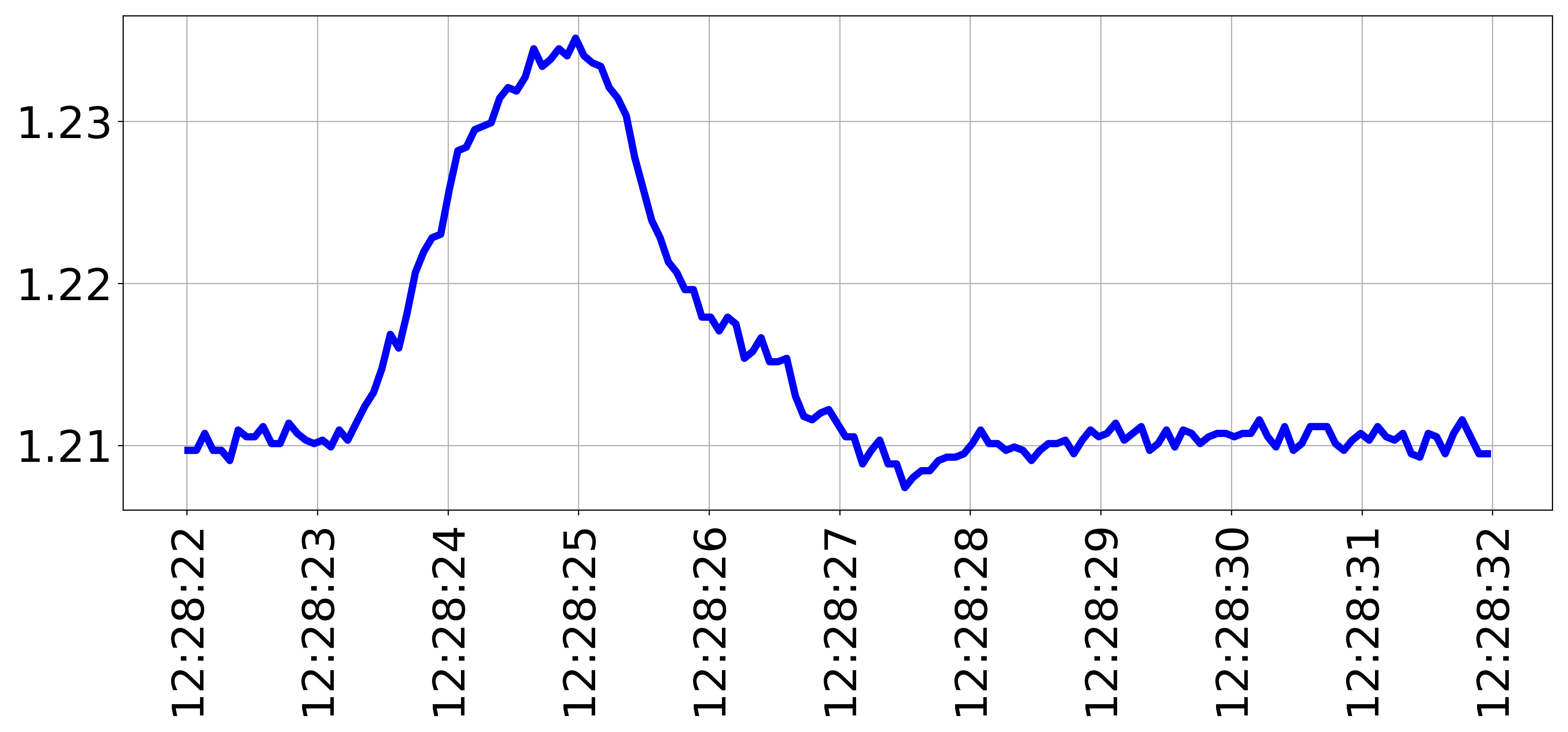}
        \end{subfigure}
        \hfill
        \begin{subfigure}[t]{0.32\textwidth}
            \includegraphics[width=\textwidth]{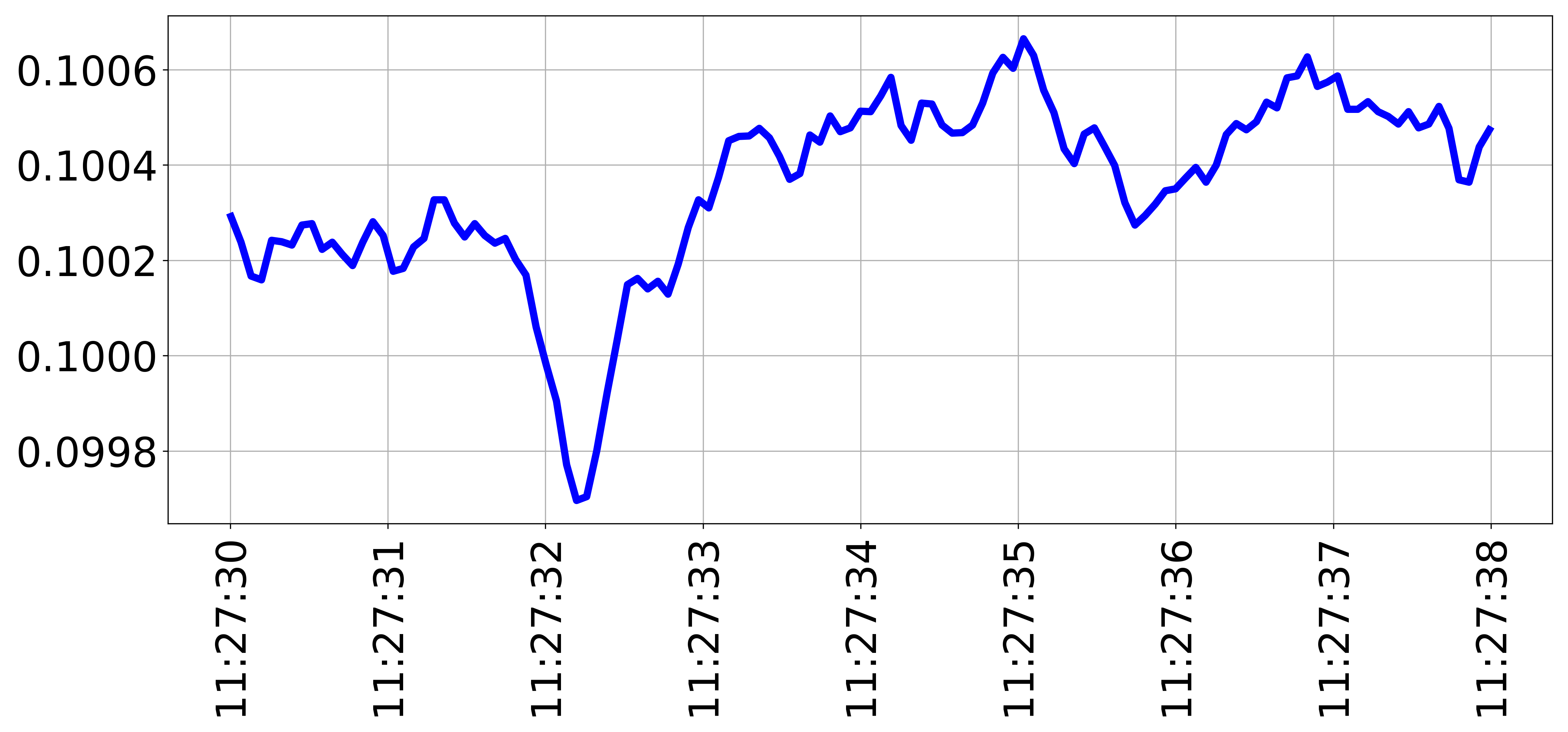}
        \end{subfigure}
        \caption*{\textbf{(a) EDA signals}}
    \end{subfigure}

    \vspace{1em}

    \begin{subfigure}[t]{\textwidth}
        \centering
        \begin{subfigure}[t]{0.32\textwidth}
            \includegraphics[width=\textwidth]{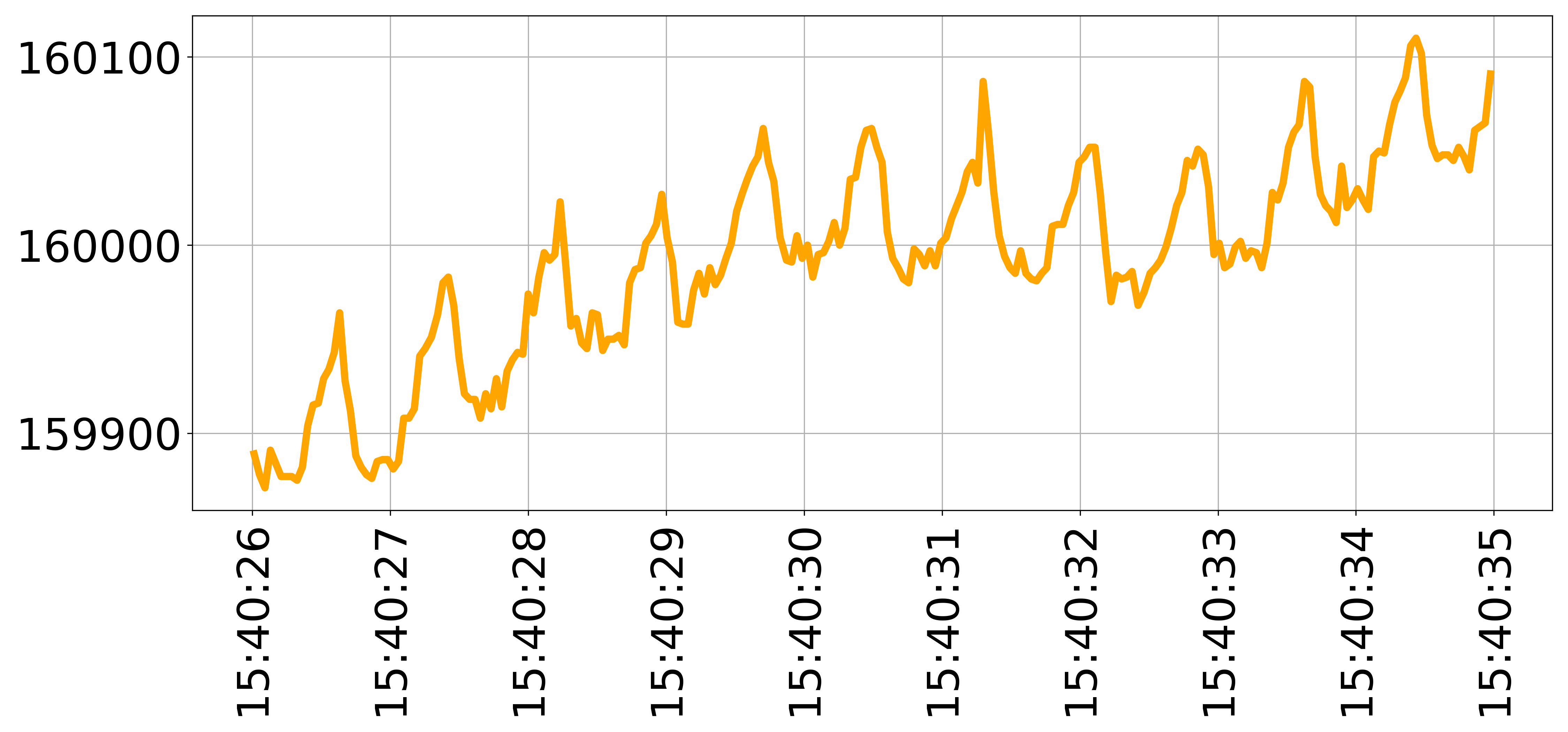}
        \end{subfigure}
        \hfill
        \begin{subfigure}[t]{0.32\textwidth}
            \includegraphics[width=\textwidth]{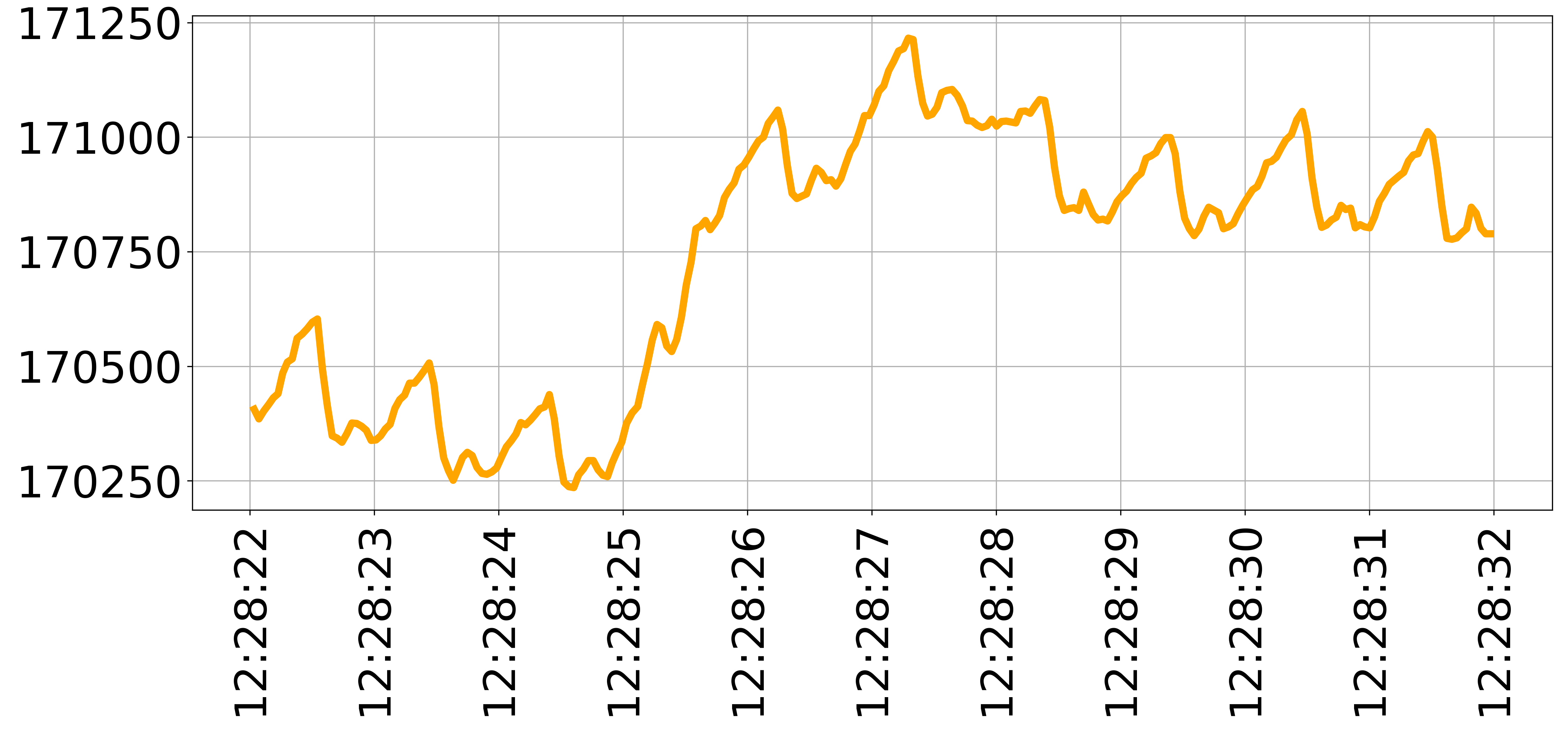}
        \end{subfigure}
        \hfill
        \begin{subfigure}[t]{0.32\textwidth}
            \includegraphics[width=\textwidth]{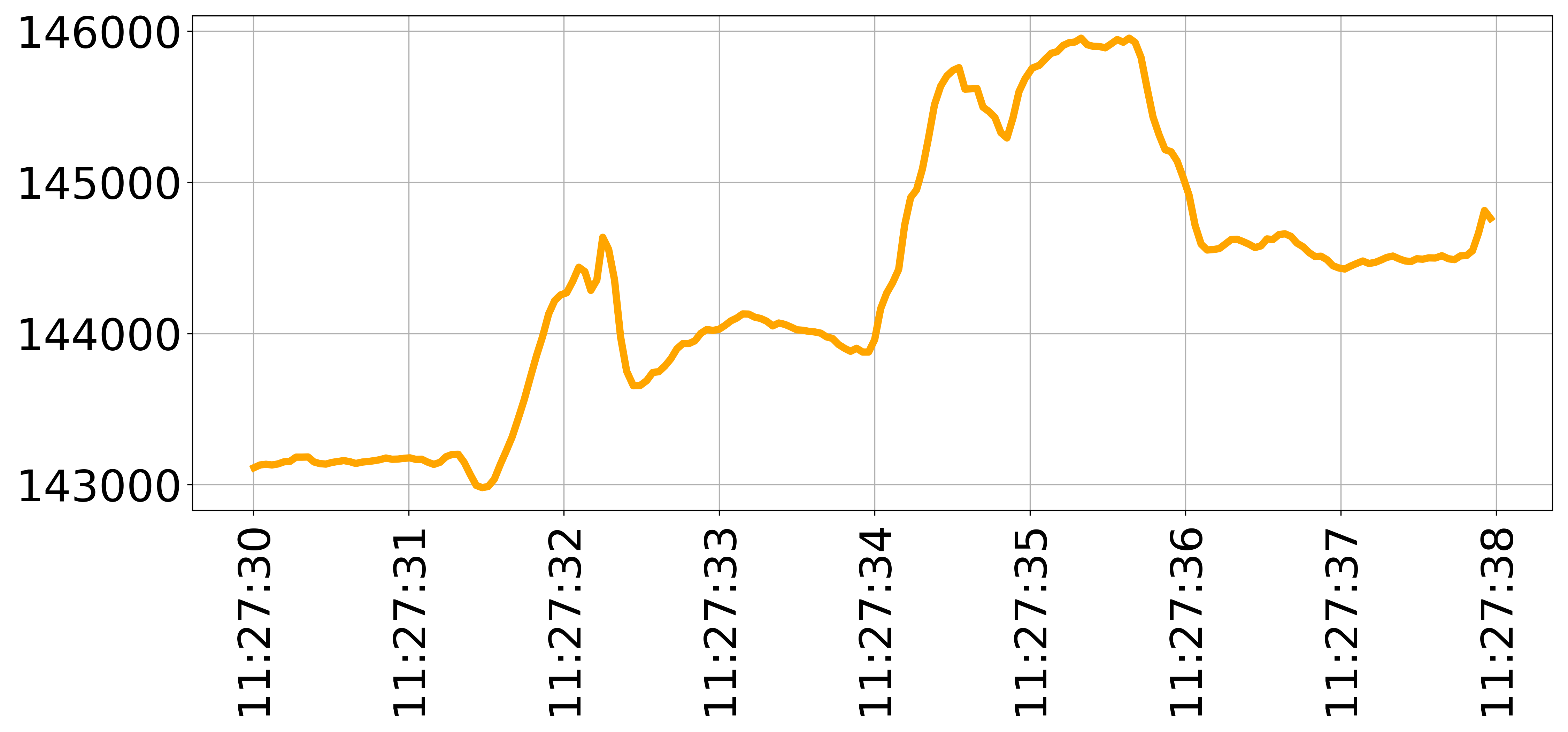}
        \end{subfigure}
        \caption*{\textbf{(b) PPG signals}}
    \end{subfigure}

    \caption{
    Visualizations of raw EDA and PPG signals for individual trials.
    Ttop row: EDA signals, Bottom row: corresponding PPG signals from the same trials. Each vertically aligned EDA–PPG pair comes from the same participant and corresponds to the same claim evaluation task. The $y$-axis indicates signal amplitude, and the $x$-axis represents time (in seconds).
    }
    \label{fig:eda_ppg_trials}
\end{figure*}

Although previous work has linked physiological responses to emotional and cognitive states \cite{bu2025dynamic, lin2023review, figner2011using, li2021sensitivity, thammasan2017familiarity, mishra2022cardiac, fazio2019repetition, hasher1977frequency, dechene2010truth, vellani2023illusory}, the application of biosignals to belief formation and truth assessment remains underexplored. Our work fills this gap by (i) presenting the first dataset focused on belief/familiarity classification from EDA and PPG signals, and (ii) analyzing the potential of these signals to capture implicit epistemic processes in misinformation exposure.

\textbf{Physiological signals in cognitive \& affective research.} Physiological signals, particularly EDA and PPG, have long been used to study affective and cognitive processes \cite{babaei2021critique, gasparini2021deep}. A growing body of literature in emotion recognition \cite{kang20221d} demonstrates that features extracted from EDA and PPG can effectively differentiate emotional states such as joy, stress, anxiety, and depression \cite{zhu2019detecting}. \citet{lin2023review} provide comprehensive overviews of how biosignals have been used to capture emotional reactions in ecologically valid, wearable settings.

Beyond affective computing, EDA and PPG have also shown promise in detecting cognitive states \cite{gasparini2021deep}, including mental workload, attention, and decision confidence. \citet{figner2011using}, for example, highlighted the role of skin conductance as an index of uncertainty and arousal during decision-making. Similarly, \citet{li2021sensitivity} demonstrated that physiological signals can serve as robust indicators of cognitive strain in dynamic tasks like driving or human–machine interaction. Despite this progress, little attention has been paid to the use of EDA or PPG in capturing epistemic states, such as belief, disbelief, or familiarity with information, especially in contexts involving repetition or misinformation.

Some prior work \cite{thammasan2017familiarity, mishra2022cardiac} has examined familiarity effects using EEG, particularly in music-related emotion studies. \citet{thammasan2017familiarity} showed that music familiarity modulates EEG spectral and connectivity patterns, and found that novel music stimuli improve emotion classification accuracy. \citet{mishra2022cardiac} further observed that familiarity during emotional encounters shapes cardiac–brain dynamics, suggesting that physiological and neural responses are tightly linked to contextual exposure. However, these studies do not address belief development, information repetition, or exposure to false claims.

Together, prior research \cite{lin2023review, figner2011using, li2021sensitivity, thammasan2017familiarity, mishra2022cardiac} has demonstrated that physiological signals reflect a range of cognitive and emotional processes. Yet the connection between involuntary physiological activity and belief or familiarity judgments, particularly in response to misinformation, remains underexplored. This gap motivates our investigation of whether EDA and PPG can serve as implicit signals of familiarity and belief when individuals assess repeated or novel information.

\textbf{Belief, repetition, and misinformation.} The formation of belief, especially in the presence of misinformation, has been extensively studied in psychology and cognitive science \cite{fazio2019repetition, hasher1977frequency, dechene2010truth, vellani2023illusory}. A well-documented phenomenon is the illusory truth effect, in which repeated exposure to a statement increases its perceived accuracy, regardless of its factual correctness. This effect persists across plausible and implausible claims, and even when individuals possess contradictory knowledge \cite{fazio2015knowledge}. Processing fluency is often cited as the underlying mechanism: repeated statements are processed more easily, which leads to a subjective sense of truth \cite{hasher1977frequency}.

Belief formation is also shaped by cognitive biases, such as motivated reasoning, source-monitoring errors, and individual traits like receptivity to ``bullshit'' or overclaiming knowledge \cite{vellani2023illusory}. These findings suggest that belief is not purely rational but is influenced by prior exposure, familiarity, and emotional context.

However, most prior studies have relied on self-report measures and behavioral responses to assess belief, which may overlook the implicit or unconscious dimensions of belief formation \cite{gao2021investigating, gao2019predictingpersonalitytraitsphysical}. A notable recent study by \citet{jiang2024repetition} found that even individuals who initially accepted climate science increased their belief in climate-skeptical claims after repeated exposure. However, their work relied solely on self-reported attitudes and did not include physiological measures.

Our study extends this line of inquiry by incorporating real-time physiological recordings in a controlled, face-to-face environment. By replicating key elements of Jiang et al.'s paradigm \cite{jiang2024repetition} and augmenting it with biosignal tracking, we investigate whether EDA and PPG can reveal involuntary markers of belief and familiarity during misinformation evaluation.

\begin{figure*}[tbp]
    \centering
    \includegraphics[trim=0.2cm 0.6cm 0.2cm 0.2cm, clip=true, width=\textwidth]{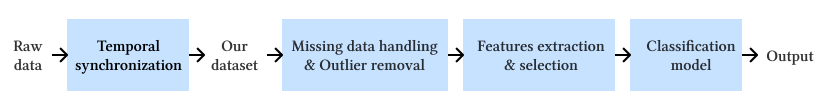}
    \caption{
    Data processing and classification pipeline. 
    }
    \label{fig:pipeline}
\end{figure*}

\textbf{Multimodal misinformation detection \& biosignals for truth evaluation.} While early misinformation studies \cite{gao2021investigating} emphasized content analysis and survey-based beliefs, recent efforts \cite{joshi2025multimodal, banik2024exploring, uskaikar2024multi} have moved toward multimodal and physiological approaches to uncover deeper cognitive and affective responses. Biosignals, such as heart rate, EDA, respiration, and facial EMG, have been widely applied in deception detection, using non-verbal cues to infer hidden or implicit states.
For example, \citet{bhamare2020deep} proposed a deep learning model that fuses multiple physiological streams with attention mechanisms to classify deceptive behavior. \citet{joshi2025multimodal} integrated EDA, keystroke patterns, and eye-tracking data to improve deception classification. Reviews such as \cite{banik2024exploring} and empirical work by \citet{uskaikar2024multi} further support the integration of central (EEG) and peripheral (EDA, ECG) signals for robust detection of cognitive conflict and emotional tension --- key elements often associated with deception or uncertainty.

However, these studies have largely focused on binary deception tasks or high-stakes lies, rather than everyday belief formation or misinformation evaluation. Notably absent is research on how physiological responses, especially EDA and PPG, vary with belief and familiarity judgments in repeated exposure scenarios. Our work addresses this gap by isolating the contributions of EDA and PPG in belief classification tasks involving true, false, repeated, and novel claims.

\textbf{Physiology-based datasets \& classification benchmarks.} Existing physiological datasets have primarily supported affective computing and stress detection research. Popular resources such as DEAP \cite{5871728, tripathi2017using, rajpoot2022subject}, WESAD \cite{10.1145/3242969.3242985, choi2025emotion, benita2024stress}, AMIGOS \cite{mirandacorrea2017amigosdatasetaffectpersonality, 10097567}, and MAHNOB-HCI \cite{10.1109/T-AFFC.2011.25, wang2023multimodal} provide EEG, EDA, and PPG data collected during emotion-evoking or stress-inducing tasks. These datasets have enabled reliable classification of valence–arousal states, mental load, and user traits like personality or mood.
However, none of these datasets directly target belief-related processes, nor do they include misinformation, repetition exposure, or truth evaluation tasks. The absence of such benchmarks has limited progress on epistemic signal classification using biosignals.

To address this, we introduce a new dataset collected in controlled, in-person sessions where participants evaluated the truthfulness of repeated and novel claims while EDA and PPG were continuously recorded. The dataset includes annotations for belief ratings, exposure history, claim accuracy, and participant response. To our knowledge, it is the first dataset designed to explore the physiological basis of belief and familiarity in a misinformation context.

\section{Dataset and Experiment}

\subsection{Our Dataset}

We introduce a novel dataset consisting of 672 labeled trials collected from 28 participants, each completing 24 claim-evaluation tasks. Throughout these tasks, EDA and PPG signals were recorded to capture involuntary physiological responses to misleading and repeated information. This dataset, to our knowledge, is the first to directly associate biosignals with belief judgment and repetition-induced familiarity in the context of misinformation.

\textbf{Setup.} The experimental design follows the three-phase illusory truth paradigm (encoding, delay, and test), as adapted from \cite{jiang2024repetition}. Participants first completed surveys assessing their attitudes toward science and climate change, including the belief in science scale and attitudes toward the scientific method.

In the encoding phase, participants viewed climate-related claims (science-backed, skeptical, and neutral) for 8 seconds each. After a 15-minute distraction task (reading and spatial rotation exercises), the test phase presented 24 claims, some repeated, some novel, in a randomized order. Participants rated the truthfulness of each claim using a 6-point Likert scale, with higher scores indicating greater perceived truth.

To ensure balanced repetition conditions, a counterbalanced design was used: each claim appeared as repeated for some participants and non-repeated for others. Each session lasted 30–45 minutes in a controlled environment without internet access, encouraging participants to rely on prior knowledge.

The dataset includes 28 university students, aged 18–34 (with average age of 23.52, standard deviation of 3.28). Of these, 83\% were non-native English speakers, 48\% identified as female, and 52\% as male. All participants consented to the study and agreed to the use of their anonymized data for research purposes.

\textbf{Biosignal collection.} Physiological signals were collected using the EmotiBit wearable sensor, worn on the left arm (see Fig. \ref{experiment_setup}). Two primary signal types were recorded: (i) EDA: Sampled at 15 Hz, reflecting sympathetic nervous system arousal, sensitive to emotional salience and novelty. (ii) PPG: Sampled at 25 Hz, providing heart rate variability metrics related to cognitive effort, uncertainty, and stress regulation.

Each participant completed 24 claim-evaluation trials, during which EDA and PPG data were continuously recorded and synchronized with the claim presentation. Additional channels (temperature, accelerometer) were used for quality control, and mouse activity and reaction time were logged for preprocessing.

\textbf{Labels \& annotations.} Each trial is annotated with the following labels (see Fig.~\ref{fig:data_align}): (i) Belief: Whether the claim was believed or not believed. (ii) Repetition: Whether the claim was repeated or novel. (iii) Combined Label: A derived four-class label combining both belief and repetition. This label reflects four distinct conditions based on the intersection of belief and repetition: Believed and Repeated (B-R), Believed and Not Repeated (B-NR), Not Believed and Repeated (NB-R), and Not Believed and Not Repeated (NB-NR). These labels support both binary (\eg, Believed \vs~ Not Believed) and multi-class classification tasks, enabling the investigation of how physiological signals relate to judgments of truth and familiarity.

Figure \ref{fig:label_dist} shows the label distribution of our dataset, while Figure \ref{fig:eda_ppg_trials} displays representative time-series plots of EDA and PPG responses.

\textbf{Why EDA and PPG?} We selected EDA and PPG due to their interpretability and non-intrusive nature, making them ideal for real-time or wearable misinformation detection systems. EDA is particularly sensitive to the emotional salience \cite{shukla2019feature, veeranki2024comparison} and novelty of stimuli, while PPG offers insights into cognitive load and decision conflict --- key components in understanding the processes underlying belief formation and the effects of repeated exposure. 

In the remainder of this paper, we focus on EDA and PPG as complementary signals for studying involuntary physiological responses to belief and repetition.
By analyzing these two signals both separately and together, our dataset provides a valuable foundation for studying involuntary indicators of belief and familiarity, paving the way for biosignal-based detection of misinformation susceptibility.

\subsection{Our Experiment}

\textbf{Data processing.} The data processing pipeline, illustrated in Figure \ref{fig:pipeline}, ensures the reliability and interpretability of physiological data before model training. First, temporal synchronization across data streams (EDA, PPG, screen recordings, and self-reports) was achieved by converting raw timestamps to absolute time values, compensating for the EmotiBit sensor’s internal clock resets.

Each participant's continuous physiological data was divided into 24 trials, representing individual claim-evaluation tasks. Timestamp markers from the screen recordings were used to segment the data (see Fig.~\ref{fig:data_align}). Missing data and outliers were handled by excluding trials with excessive artifacts and interpolating signal gaps. To minimize inter-individual variability, data was normalized within each participant’s trial using z-scores.

Feature extraction followed a structured approach, with EDA and PPG data analyzed separately. Features were organized into three categories: (i) statistical and time-domain features (\eg, mean, median, RMS, skewness), (ii) frequency-domain features (\eg, power spectral density, peak frequency, spectral entropy), and (iii) complexity measures (\eg, Shannon entropy, Hurst exponent, DFA). These features capture the temporal dynamics and non-linear patterns in physiological responses.

To improve model performance and reduce dimensionality, we used a feature selection process, inspired by \citet{pohjalainen2015feature}, which ranked features based on their discriminative capability.

\textbf{Classification models.} We adopted three machine learning models to assess the potential of physiological signals in classifying belief and familiarity: KNN, CNN, and LightGBM. These models were selected to benchmark different approaches, including traditional feature-based methods (KNN, LightGBM) and deep learning (CNN), to explore the efficacy of each strategy for biosignal classification.

\textbf{Justification for model selection.} KNN was chosen as a baseline model for its simplicity and interpretability. It is effective for early-stage experimentation, particularly when working with handcrafted features, and it is sensitive to the local structure of data. Its inclusion in the feature selection process, as outlined by \cite{pohjalainen2015feature}, enables the identification of the most relevant features while offering clear insights into the structure of the dataset. 

LightGBM was selected due to its ability to capture non-linear relationships and handle feature interactions, making it ideal for complex datasets like physiological signals. As a tree-based model, it is capable of efficiently processing high-dimensional data while providing robustness against overfitting. LightGBM’s use in emotion recognition \cite{10698721} justifies its selection for tasks like belief and familiarity classification. 

CNN was included as a deep learning approach, offering a complementary evaluation to feature-based models. CNNs excel at automatic feature extraction and are well-suited to time-series data, such as physiological signals \cite{rajpoot2022subject, benita2024stress}, where temporal dependencies are important \cite{wang2023robust,wang2024taylor}. The inclusion of CNN enables us to benchmark the performance of deep learning models, which can potentially capture more complex, hierarchical patterns in the raw data.

\textbf{Metrics.} Each model was trained and evaluated separately for EDA and PPG features, without combining signal types, to assess their distinct contributions. We investigated three classification tasks:
\renewcommand{\labelenumi}{\roman{enumi}.}
\begin{enumerate}[leftmargin=0.6cm]
\item Belief classification (binary): Believe \vs~ Not Believe
\item Repetition classification (binary): Repeated \vs~ Not Repeated
\item four-class classification: Four-class label combining belief and repetition patterns
\end{enumerate}
A participant-wise data split was used, with 20\% reserved for testing and 80\% for training, ensuring generalizability across individuals. Evaluation metrics included accuracy, macro precision, macro recall, macro F1 score, and confusion matrix.

\textbf{Why use macro precision, recall, and F1 score?} Macro precision, recall, and F1 score are used to evaluate classification performance in multi-class tasks, especially when class imbalances exist (see Fig.~\ref{fig:label_dist}). These metrics compute the performance for each class independently and then average the results, ensuring equal weight for all classes, regardless of their frequency. This approach provides a more balanced evaluation of the model's ability to correctly predict each class, offering a clearer picture of its overall performance, particularly in our context of physiological signal-based classification tasks.

\textbf{Evaluation.} To optimize model performance, we used the feature selection framework from \citet{pohjalainen2015feature}, which applies KNN in an inner cross-validation loop to assess feature ranking and selection strategies. We experimented with feature sets ranging from 7 to 15 top-ranked features, selecting the most discriminative features for each classification task. This allowed us to identify the key features that contributed to the models’ ability to distinguish between belief, repetition, and their combination.

By comparing KNN, CNN, and LightGBM, we aimed to evaluate different modeling strategies and assess the effectiveness of EDA and PPG signals in predicting belief and familiarity. 
This multi-model approach provides insights into the relationship between physiological responses and misinformation susceptibility, as well as the strengths and limitations of each modeling technique.

\section{Results and Discussion}

\subsection{Comparison}

We evaluated three classification tasks: belief classification, repetition classification, and combined belief-repetition group classification, using KNN, LightGBM, and CNN across two physiological signal types (EDA and PPG). Each model was assessed using four standard metrics: Accuracy, Macro Precision, Macro Recall, and Macro F1 Score. The findings reveal nuanced patterns across tasks and signal types:

\begin{table}[tbp]
    \centering
    \renewcommand{\arraystretch}{1.3}
    \begin{tabularx}{\columnwidth}{l l >{\centering\arraybackslash}X >{\centering\arraybackslash}X}
        \toprule
        \textbf{Model} & \textbf{Metrics} & \textbf{EDA} & \textbf{PPG} \\
        \midrule
        \multirow{4}{*}{\textbf{KNN}}
            & Accuracy         & 67.83 & 59.72 \\
            & Precision  & 77.12 & 60.31 \\
            & Recall     & 63.28 & 54.60 \\
            & F1 Score   & 60.77 & 49.58 \\
        \midrule
        \multirow{4}{*}{\textbf{LightGBM}}
            & Accuracy         & 59.72 & 61.90 \\
            & Precision  & 63.49 & 54.18 \\
            & Recall     & 61.88 & 52.76 \\
            & F1 Score   & 58.99 & 51.32 \\
        \midrule
        \multirow{4}{*}{\textbf{CNN}}
            & Accuracy         & 63.43 & 62.22 \\
            & Precision  & 59.81 & 56.39 \\
            & Recall     & 56.87 & 51.16 \\
            & F1 Score   & 56.06 & 43.39 \\
        \bottomrule
    \end{tabularx}
    \caption{
    \textit{Belief classification} performance comparison of KNN, LightGBM, and CNN using EDA and PPG. Evaluation metrics include Accuracy, Macro-Precision, Macro-Recall, and Macro-F1 Score (all reported as percentages). The optimal $K$ values for the KNN model, determined via cross-validation, were $K = 26$ for EDA and $K = 19$ for PPG.}
    \label{tab:belief_all_models}
\end{table}

\textbf{Belief classification.} KNN consistently outperforms LightGBM and CNN in belief classification using EDA (see Table~\ref{tab:belief_all_models}). It achieves the highest scores across nearly all metrics, including Accuracy (67.83\%), Macro Precision (77.12\%), and Macro F1 Score (60.77\%), suggesting that EDA carries discriminative signals reflecting belief judgments. For PPG, CNN obtains the highest Accuracy (62.22\%), but its F1 Score (43.39\%) trails behind KNN (49.58\%), indicating that its predictions are less balanced across classes.

The relatively high precision and moderate recall for KNN suggest it is conservative but reliable in belief detection using EDA, whereas PPG may be more sensitive to model choice and less informative on its own.

\begin{table}[tbp]
    \centering
    \renewcommand{\arraystretch}{1.3}
    \begin{tabularx}{\columnwidth}{l l >{\centering\arraybackslash}X >{\centering\arraybackslash}X}
        \toprule
        \textbf{Model} & \textbf{Metrics} & \textbf{EDA} & \textbf{PPG} \\
        \midrule
        \multirow{4}{*}{\textbf{KNN}}
            & Accuracy        & 63.64 & 65.97 \\
            & Precision & 64.15 & 66.05 \\
            & Recall    & 63.57 & 65.97 \\
            & F1 Score  & 63.20 & 65.92 \\
        \midrule
        \multirow{4}{*}{\textbf{LightGBM}}
            & Accuracy        & 67.13 & 63.19 \\
            & Precision & 67.42 & 63.33 \\
            & Recall    & 67.18 & 63.19 \\
            & F1 Score  & 67.01 & 63.10 \\
        \midrule
        \multirow{4}{*}{\textbf{CNN}}
            & Accuracy        & 57.46 & 54.07 \\
            & Precision & 67.74 & 58.59 \\
            & Recall    & 57.46 & 54.34 \\
            & F1 Score  & 42.39 & 47.67 \\
        \bottomrule
    \end{tabularx}
    \caption{
    \textit{Repetition classification} performance comparison of KNN, LightGBM, and CNN using EDA and PPG. Optimal values of $K$ for the KNN model were selected via cross-validation: $K = 31$ for EDA and $K = 13$ for PPG.}
    \label{tab:repetition_all_models}
\end{table}

\begin{figure*}[tbp]
    \centering
    \begin{subfigure}{0.16\textwidth}
        \includegraphics[trim=3cm 2.5cm 0cm 0cm, clip=true, width=\linewidth]{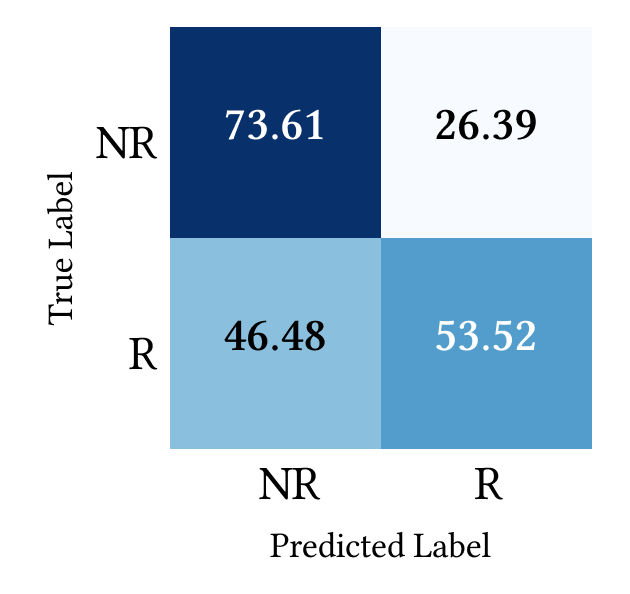}
        \caption{EDA - KNN}
        \label{fig:repetition_eda_knn}
    \end{subfigure}
    \begin{subfigure}{0.16\textwidth}
        \includegraphics[trim=3cm 2.5cm 0cm 0cm, clip=true, width=\linewidth]{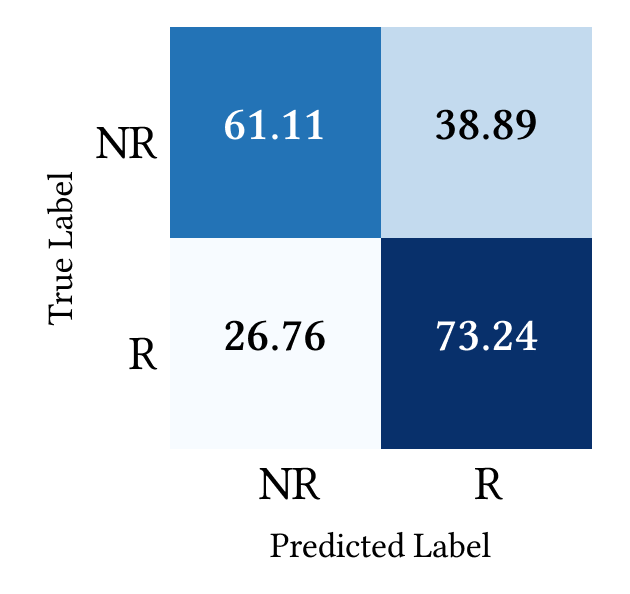}
        \caption{EDA - LightGBM}
        \label{fig:repetition_eda_lightgbm}
    \end{subfigure}
    \begin{subfigure}{0.16\textwidth}
        \includegraphics[trim=3cm 2.5cm 0cm 0cm, clip=true, width=\linewidth]{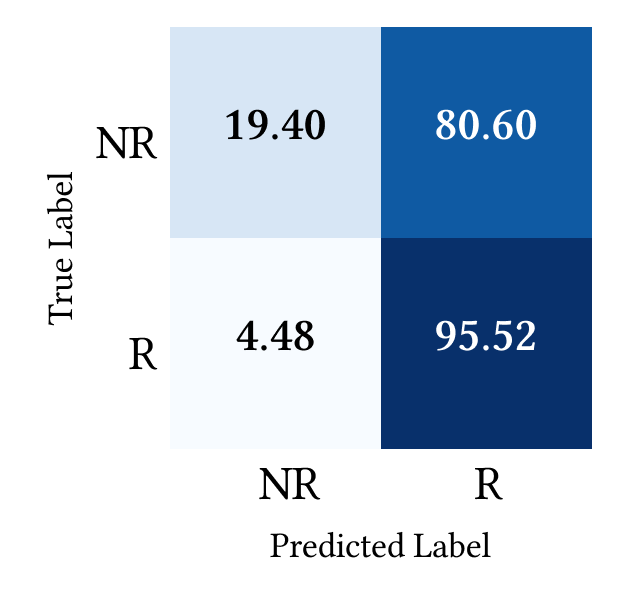}
        \caption{EDA - CNN}
        \label{fig:repetition_eda_cnn}
    \end{subfigure}
    \hspace{0.5em}
    \begin{subfigure}{0.16\textwidth}
        \includegraphics[trim=3cm 2.5cm 0cm 0cm, clip=true, width=\linewidth]{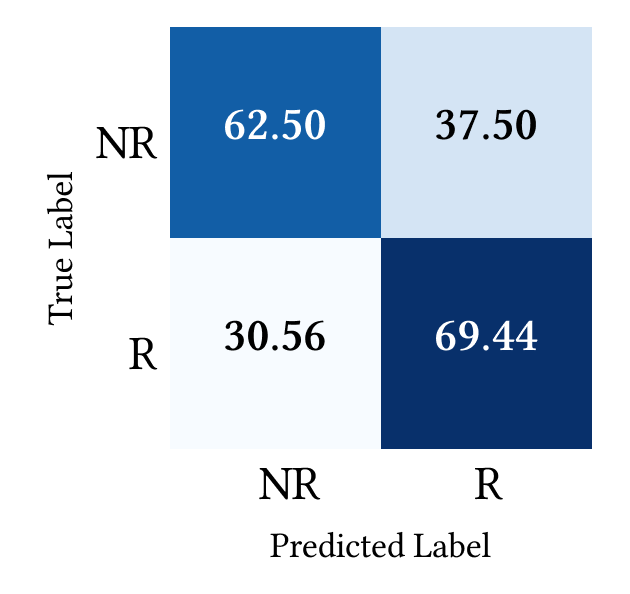}
        \caption{PPG - KNN}
        \label{fig:repetition_ppg_knn}
    \end{subfigure}
    \begin{subfigure}{0.16\textwidth}
        \includegraphics[trim=3cm 2.5cm 0cm 0cm, clip=true, width=\linewidth]{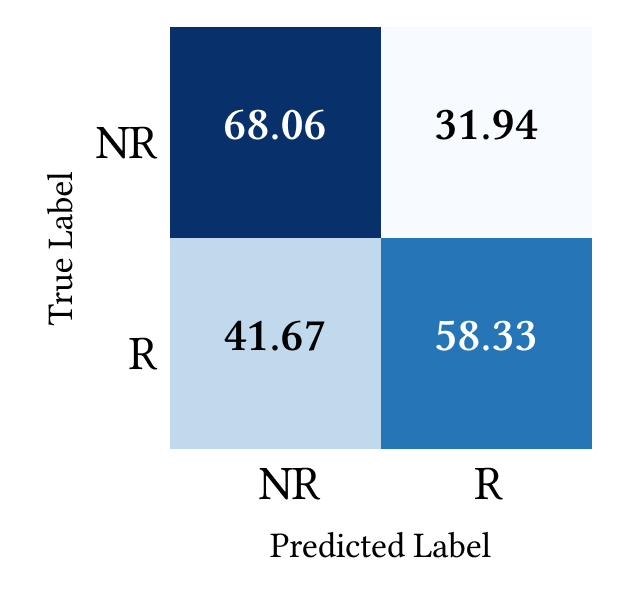}
        \caption{PPG - LightGBM}
        \label{fig:repetition_ppg_lightgbm}
    \end{subfigure}
    \begin{subfigure}{0.16\textwidth}
        \includegraphics[trim=3cm 2.5cm 0cm 0cm, clip=true, width=\linewidth]{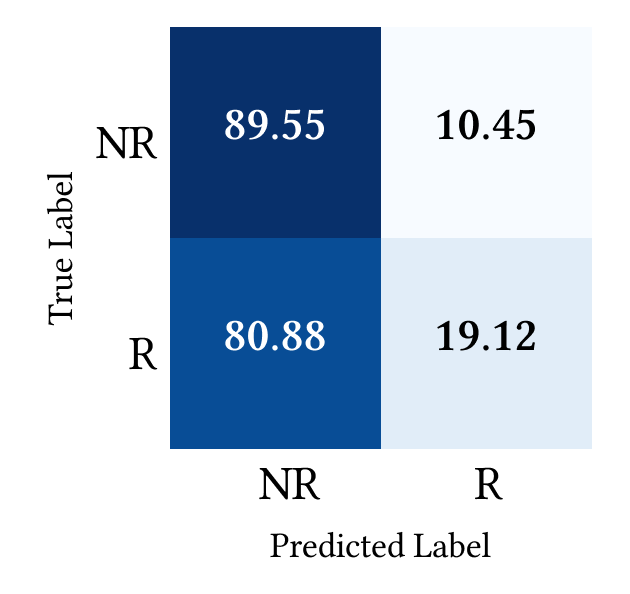}
        \caption{PPG - CNN}
        \label{fig:repetition_ppg_cnn}
    \end{subfigure}
    \caption{Confusion matrices for \textit{repetition classification} using KNN, LightGBM, CNN. Subfigures (a–c) display the classification results using EDA features, while subfigures (d–f) show the results using PPG features. In each matrix, the horizontal axis represents the predicted labels, and the vertical axis represents the ground truth labels.}
    \label{fig:confusion_matrices_repetition}
\end{figure*}

\textbf{Repetition classification.} As shown in Table~\ref{tab:repetition_all_models}, LightGBM on EDA emerges as the top performer, achieving the highest scores across all metrics, including Accuracy (67.13\%) and Macro F1 Score (67.01\%). This indicates that tree-based models effectively capture the physiological signal patterns associated with repetition effects. Interestingly, for PPG, KNN performs best, reaching Accuracy (65.97\%) and Macro F1 Score (65.92\%), suggesting complementary modality-model dynamics.

This divergence highlights a possible signal-specific affinity: tree-based models may be better suited to the relatively stable patterns in EDA for repetition, while instance-based models like KNN can adapt better to variability in PPG.

Figure \ref{fig:confusion_matrices_repetition} presents confusion matrices.
As illustrated, the classification performance varies notably across models and signal types, reflecting the inherent complexity of our dataset. These results highlight the challenge of inferring repetition-related cognitive states from biosignals and underscore the potential of our dataset as a new avenue for exploring the physiological correlates of misinformation processing.

\textbf{Four-class classification (Belief + Repetition).} Table~\ref{tab:groups_all_models} summarizes the results. This four-class task is notably more challenging, with all models showing reduced performance. The best result is achieved by KNN on EDA, with Accuracy at 45.45\% and Macro F1 Score at 32.84\%. Performance across both signal types and all models is modest, with CNN again lagging, particularly in terms of F1 scores (13.39\% on EDA, 13.32\% on PPG), suggesting difficulties in generalizing over compound belief-repetition labels.

The significant performance drop indicates increased task complexity and possible overlapping signal patterns when belief and repetition are combined. It also suggests that raw physiological features may not linearly encode such multi-faceted cognitive states, potentially requiring multimodal fusion or hierarchical modeling.

\begin{table}[tbp]
    \centering
    \renewcommand{\arraystretch}{1.3}
    \begin{tabularx}{\columnwidth}{l l >{\centering\arraybackslash}X >{\centering\arraybackslash}X}
        \toprule
        \textbf{Model} & \textbf{Metrics} & \textbf{EDA} & \textbf{PPG} \\
        \midrule
        \multirow{4}{*}{\textbf{KNN}}
            & Accuracy        & 45.45 & 37.50 \\
            & Precision & 33.16 & 24.69 \\
            & Recall    & 37.54 & 29.85\\
            & F1 Score  & 32.84 & 25.09\\
        \midrule
        \multirow{4}{*}{\textbf{LightGBM}}
            & Accuracy        & 42.36 & 38.89 \\
            & Precision & 40.82 & 27.56 \\
            & Recall    & 34.29 & 31.33 \\
            & F1 Score  & 30.83 & 27.55 \\
        \midrule
        \multirow{4}{*}{\textbf{CNN}}
            & Accuracy        & 36.57 & 36.30 \\
            & Precision & 9.21  & 9.14 \\
            & Recall    & 24.50 & 24.50 \\
            & F1 Score  & 13.39 & 13.32 \\
        \bottomrule
    \end{tabularx}
    \caption{
    \textit{Four class task} performance comparison of KNN, LightGBM, and CNN using EDA and PPG. This multi-class task involves predicting one of four distinct user response conditions: 
        \textit{Believed and Repeated (B-R)}, \textit{Believed and Not Repeated (B-NR)}, \textit{Not Believed and Repeated (NB-R)}, and \textit{Not Believed and Not Repeated (NB-NR)}. 
        For the KNN model, the best-performing $K$ values were $K = 40$ for EDA and $K = 47$ for PPG, selected via cross-validation.
        }
    \label{tab:groups_all_models}
\end{table}





\subsection{Discussion}

Our results offer several compelling insights into how physiological signals, specifically EDA and PPG, can serve as indicators of belief, repetition, and their interaction. The findings reveal not only the discriminative potential of these signals but also the nuanced relationship between signal type, task complexity, and model behavior.

\textbf{Physiological signals: EDA \vs~PPG.} A clear trend emerges across all tasks: EDA consistently yields higher classification performance than PPG. This suggests that electrodermal activity is more tightly coupled with the cognitive and emotional underpinnings of belief and repetition than photoplethysmography signals, which primarily capture cardiovascular dynamics. EDA’s stronger performance likely reflects its sensitivity to arousal and attentional states, which have been shown to modulate belief updating and familiarity effects.

Interestingly, PPG's role is less consistent. While it performs reasonably well in some settings (\eg, repetition classification with KNN), it may require deeper temporal modeling or signal decomposition to unlock its full potential \cite{wang2023robust,wang2024meet}. This opens up a promising direction: multi-modal fusion, where the complementary nature of EDA and PPG could be used to capture both autonomic and cognitive-emotional cues \cite{ding2025learnable}.

\textbf{Model behavior and task alignment.} Among the models, KNN demonstrates surprising robustness, especially in belief classification and the more complex four-class task. Its non-parametric nature and sensitivity to local patterns may explain its advantage when dealing with high intra-subject variability or when the feature space lacks strong global structure.

In contrast, LightGBM shines in repetition classification, suggesting that decision trees can effectively partition the physiological space based on repetition-induced signal patterns. This might reflect more regular or predictable physiological changes during repeated exposure, which decision trees can exploit.

CNNs underperform across the board, particularly in the multi-class group task. This may be due to the relatively shallow architecture used and the lack of large-scale data to effectively learn discriminative representations. CNNs often require more complex temporal encoding \cite{wang2024meet} or spatial augmentation \cite{wang2024adaptive} to outperform classical models, especially when data is limited.

\textbf{Task complexity and signal discriminability.} As task complexity increases, from binary belief classification to four-class belief-repetition group classification, performance drops substantially. This is expected, but it also highlights two key challenges:
\renewcommand{\labelenumi}{\roman{enumi}.}
\begin{enumerate}[leftmargin=0.6cm]
\item Physiological signals may not encode complex cognitive states in a linearly separable fashion.
\item There may be latent interdependencies between belief and repetition that simple models cannot capture.
\end{enumerate}
These findings motivate the development of hierarchical, multi-task, or representation learning approaches that can disentangle shared and unique physiological signatures associated with belief and repetition.

\textbf{Implications for trust-aware HCI.} From a human-computer interaction (HCI) perspective, the results suggest that involuntary physiological signals can provide implicit cues about user belief states and information familiarity. This opens avenues for adaptive systems that monitor trustworthiness perceptions or detect potential misinformation influence in real-time, without requiring explicit user feedback.

However, system designers must contend with signal ambiguity, inter-individual variability, and context-dependence. Future systems should therefore incorporate personalized baselines, contextual calibration, or meta-learning techniques to accommodate these complexities.

\textbf{Understanding the role of signal type and data characteristics.} Our results underscore the importance of both signal type and dataset characteristics in physiological classification tasks. Across all three tasks, belief, repetition, and group classification, EDA consistently outperformed PPG in F1 score and precision. This supports existing literature that highlights EDA’s strong sensitivity to emotional arousal and attentional modulation, two processes likely engaged when individuals assess the truthfulness or familiarity of information. In contrast, PPG, which reflects cardiovascular dynamics, may be less directly linked to cognitive states like belief and is more prone to noise from movement or posture changes. These findings suggest that EDA may be the more effective and robust signal type for modeling internal states such as belief, trust, and uncertainty, particularly in dynamic or real-world human-computer interaction scenarios.

In addition to signal type, the composition and distribution of our dataset played a crucial role in shaping model performance. While the repetition classification task had nearly balanced labels, the belief task exhibited a moderate skew toward the ``Believed'' class (62.15\%). This class imbalance may have influenced models, especially KNN, to favor the majority class, resulting in reduced recall and F1 scores for the minority ``Not Believed'' class. The effect was even more pronounced in the four-class task, which involved four distinct label combinations based on belief and repetition. Certain user states, particularly ``Not Believed and Repeated'' (NB-R), were substantially underrepresented, contributing to poor macro F1 performance across all models. This highlights a common but under-addressed challenge in biosignal-based modeling: the difficulty of learning from rare yet cognitively meaningful states. Future work should explore class rebalancing methods such as oversampling, synthetic augmentation (\eg, SMOTE \cite{fernandez2018smote, yoshikawa2021time, reddy2024enhancing}), or class-weighted loss functions to improve robustness and fairness.

Finally, our findings demonstrate that low-cost, non-invasive physiological signal types like EDA and PPG can provide valuable insights into belief and repetition responses, traditionally measured using more intrusive methods such as EEG. This opens new possibilities for designing scalable, real-time systems that detect user trust, skepticism, or familiarity in misinformation contexts without requiring laboratory-grade equipment. By bridging signal type, dataset dynamics, and real-world applicability, this study provides a promising foundation for future work in trust-aware and cognition-aware interfaces.

\subsection{Future Directions}


Building on this study, several promising research avenues emerge that could enhance both the performance and applicability of physiological signal-based belief and repetition classification. One compelling direction lies in multimodal learning \cite{ding2025learnable}. While EDA and PPG provide valuable insights into autonomic responses, they represent only a fraction of the human physiological and cognitive spectrum. Integrating these signals with additional modalities, such as eye-tracking, facial expressions, or EEG, could offer a more holistic understanding of the involuntary processes underlying belief formation and familiarity. This fusion could unlock complementary cues that are difficult to capture through any single signal type, improving model robustness and contextual sensitivity.

Another critical path forward involves temporal modeling. The current study relies on handcrafted features and shallow models, which may overlook the temporal dynamics inherent in physiological responses. Sequential architectures such as LSTMs, temporal convolutional networks, or Transformers are well-suited to capture evolving patterns over time, particularly in tasks like belief trajectory analysis or sustained attention monitoring. These architectures could be especially beneficial in tracking gradual shifts in belief states as users engage with repeated or conflicting information.

In addition, representation learning techniques, especially self-supervised or contrastive learning methods, hold great promise. Such approaches can uncover latent structure in physiological data without requiring extensive labels, which is particularly advantageous given the cost and variability of biosignal collection. By learning generalizable representations, models can better handle inter-individual differences and adapt to novel contexts or unseen misinformation content.

Given the high variability in physiological responses across individuals, there is a growing need for personalized modeling approaches. This could involve calibrating models to an individual’s baseline physiological profile, using meta-learning to rapidly adapt to new users, or incorporating contextual information to better interpret ambiguous signals. Personalization is not only key to improving accuracy but also crucial for the ethical deployment of trust-aware systems that are sensitive to user differences.

Lastly, real-world deployment and evaluation of these models will be essential to assess their practical value. Future work should explore how belief and familiarity prediction performs in ecologically valid settings, such as during social media use, news browsing, or interactive educational systems. These scenarios introduce new challenges, including noise, multitasking, and naturalistic distractions, but also provide the ultimate testbed for trust-aware human-computer interaction systems grounded in physiological signals.

\section{Conclusion}

This study investigates the potential of physiological signals, specifically electrodermal activity (EDA) and photoplethysmography (PPG) to serve as involuntary indicators of belief and familiarity in response to misinformation, with an emphasis on repetition-based effects. Through a novel, in-person misinformation evaluation task, we collected a unique dataset and developed a synchronized pipeline for reliable feature extraction and analysis.

We benchmarked three modeling approaches, KNN, LightGBM, and CNN, representing traditional machine learning and deep learning paradigms. Our results demonstrate that physiological signals contain meaningful patterns that distinguish between belief states and exposure to repetition, with varying effectiveness across models and tasks.
By linking biosignal dynamics to implicit cognitive responses, our work offers a new perspective on how belief and familiarity are encoded physiologically. This lays the foundation for future research in real-time misinformation detection, multi-modal signal integration, and adaptive human-centered interventions.

\bibliographystyle{ACM-Reference-Format}
\bibliography{references} 
\end{document}